\documentclass[a4paper, 10 pt, conference]{ieeeconf}  
\IEEEoverridecommandlockouts
\usepackage{graphicx}      
\usepackage{grffile}       
\usepackage{booktabs}       
\usepackage{amsmath}
\usepackage{mathtools}
\usepackage{amssymb}
\usepackage{hyperref}
\usepackage{tikz}
\usepackage[math]{blindtext}
\usepackage{tikzscale}
\usepackage{bm}
\usepackage{ifthen}
\usepackage[ruled,noend]{algorithm2e}
\SetKwRepeat{Do}{do}{while}
\usepackage[author={Accacio}]{pdfcomment}

\newif\ifdebug%

\usepackage[acronym]{glossaries}%

\newcommand{\acrSing}[3]{\newacronym{#1}{#2}{#3}
  \expandafter\newcommand\csname #1\endcsname{\gls{#1}}}

\newcommand{\acrPl}[5]{
  \newacronym[plural=#4,firstplural=#5 (#4)]{#1}{#2}{#3}
  \expandafter\newcommand\csname #1\endcsname{\gls{#1}}
  \expandafter\newcommand\csname #4\endcsname{\glspl{#1}}
}

\usepackage{color}
\newcommand{\no}[1]{}

\newcommand{\final}{\debugfalse}

\newtheorem{problem}{Problem}
\newtheorem{remark}{Remark}
\newtheorem{assumption}{Assumption}

\newcommand{\RR}{\mathbb{R}}

\newcommand{\T}{^{\mathrm{T}}}
\newcommand{\1}{\mathbf{1}}

\newcommand{\norm}[1]{\|#1\|}

\newcommand{\p}{^{(p)}}
\newcommand{\pplusone}{^{(p+1)}}
\renewcommand{\vec}[1]{\bm{#1}}
\newcommand{\enstq}[2]{\{#1\mathrel{}\mid\mathrel{}#2\}}

\graphicspath{{img/}}

\usepackage{geometry}
\usepackage{stfloats}
\title{\LARGE \bf
Detection and Mitigation of Corrupted Information in Distributed Model Predictive Control Based on Resource Allocation
}

\author{Rafael Accácio Nogueira, Romain Bourdais and Hervé Guéguen
\thanks{The authors are with IETR-CentraleSupélec, 35510\newline Cesson-Sévigné, Ille-et-Vilaine, France \newline {\tt\small
\{rafael-accacio.nogueira, romain.bourdais, herve.gueguen\}
@centralesupelec.fr}\newline
}%
}

\makeatletter
\hypersetup{
    bookmarks=true,         
    unicode=false,          
    pdftoolbar=true,        
    pdfmenubar=true,        
    pdffitwindow=false,     
    pdfstartview={FitH},    
    pdftitle={\@title},    
    pdfauthor={\@author},     
    pdfsubject={Automatique},   
    pdfcreator={Rafael Accácio},   
    pdfkeywords={keyword1, key2, key3}, 
    pdfnewwindow=true,                  
    colorlinks=true,            
    linkcolor=black, 
    citecolor=black,          
    filecolor=black,          
    urlcolor=black            
}
\makeatother

\final%

\pubid{\begin{minipage}{\textwidth}\ \\[50pt]\centering
    \copyright{} 2021 IEEE.
Personal use of this material is permitted.  Permission from IEEE must be obtained for all other uses, in any current or future media, including reprinting/republishing this material for advertising or promotional purposes, creating new collective works, for resale or redistribution to servers or lists, or reuse of any copyrighted component of this work in other works.
\end{minipage}}
\begin{document}
\maketitle

\begin{abstract}
  In distributed predictive control structures, communication among agents is required to achieve a consensus and approach an optimal global behavior. Such negotiation mechanisms are sensitive to attacks on these exchanges.
  This paper proposes a monitoring scheme that detects and mitigates these attacks' effects in a resource allocation framework.
  The performance of the proposed method is illustrated through simulations of the temperature control of multiple rooms under power scarcity.
\end{abstract}

\section{INTRODUCTION}
Recent performance objectives require systems to be driven not in isolation but
in a coordinated way, emphasizing large systems.
These systems cover many applications, such as energy distribution systems,
traffic management in Smart City environments, coordinated control of
intelligent building systems, and many others.
Many works are built around model predictive control~\cite{CamachoBordons2007} to integrate optimality and constraints.

Furthermore, distributed model predictive control (DMPC)~\cite{MaestreEtAl2014}
techniques are a promising way to handle the optimization problem's complexity.
In these structures, there is no longer a single controller for all systems.
Instead, we use a set of local communicating controllers.
These strategies thus reduce the computing burden while increasing
confidentiality.

Many works use distributed optimization techniques, such as
Lagrangian relaxation~\cite{BourdaisEtAl2012}, Alternating Direction Method of
Multipliers (ADMM)~\cite{BoydEtAl2011}, primal
decomposition~\cite{PaulenEtAl2016}, dual decomposition~\cite{VelardeEtAl2017,VelardeEtAl2017a,PflaumEtAl2014}, and others~\cite{ChanfreutEtAl2018,Cohen1978}.
In these methods, local agents interact with a coordinator who uses an iterative process to ensure convergence towards the solution of the initial
problem.

Usually, it is assumed that all agents work in perfect cooperation.
However, when it is not the case, these uncooperative behavior have pernicious effects on the overall system, and their impact can be studied.
The cause of this disruptive behavior can be either involuntary due to hacking or malfunctioning, or voluntary, by developing selfish behavior.
Recent work has begun to explore this issue.
In the article~\cite{VelardeEtAl2017}, the authors are interested in the vulnerabilities induced when distributed predictive control is built on dual decomposition.
They analyze the impact of the deception depending on where it occurs: either in the followed references or directly in the local cost functions or coupling constraints.
The same authors propose defense strategies against these attacks, either by using secure scenarios based on reliable historical data~\cite{VelardeEtAl2017a} or by ignoring extreme values of control signals~\cite{VelardeEtAl2017}.
Then~\cite{ChanfreutEtAl2018} extends the initial work to analyze the vulnerabilities of the Jacobi-Gauss decomposition method.

Another way of dealing with these changes in behavior can be using robust distributed control principles, coupled with hierarchical identification of the attack~\cite{BraunEtAl2020}, or the introduction of probabilistic models that implement a resilient strategy if the information exchanged is outside the confidence interval~\cite{AnandutaEtAl2020}.

In this work, we analyze the exchange among agents controlled by DMPC using primal decomposition, which is perfectly adapted for agents that share resources.
More specifically, we are interested when malicious agent steers these exchanges.
By exploiting the nominal structure that characterizes the communication between the agents and the coordinator, we propose a monitoring scheme that detects an attack, and if necessary, corrects it.

The remainder of this paper is organized as follows.
First, in Section~\ref{sec:PS}, the primal decomposition-based DMPC is introduced.
In Section~\ref{sec:attack}, we discuss a model of the agents' selfish behavior that exploits the vulnerabilities of this DMPC structure.
Then, in Section~\ref{sec:mitigation}, we discuss the structure of the DMPC and how we can exploit it to construct a defense scheme to counteract the selfish agent.
At last, we present a particular mechanism to detect the agents' selfish behavior and mitigate its effects.
Moreover, in Section~\ref{sec:App_dMPC}, an application is given to illustrate and evaluate the algorithm's performance.
Finally, in Section~\ref{sec:CC}, we conclude, and we give an outlook of future works.

\section{PRELIMINARIES AND PROBLEM STATEMENT}\label{sec:PS}
\emph{Notation:} In this paper, $\norm{\cdot}$ and $\norm{\cdot}_{F}$ represent the $\ell_{2}$ and Frobenius norms. $\norm{\vec{v}}_{Y}$ is the weighted norm, $\norm{Y^{\frac{1}{2}}\vec{v}}$.
$P_{\mathsf{T}}(\cdot)$ is the Euclidean projection onto set $\mathsf{T}$.
$\otimes$ represents the Kronecker product. $\1_{m,n}$ is a ${m\times n}$ matrix filled with $1$.
$I_{c}$ is a ${c\times c}$ identity matrix.
$\pi_{v}$ denotes the number of elements in $v$.
A vector $\vec{v}_{i}$, correspond to the $i$-th agent, and these vectors can be stacked in a vector $\vec{v}$.

\subsection{Model Predictive Control}\label{ssec:MPC}
Our primary purpose is to control a system composed of $M$ subsystems using MPC\@.
The dynamics of the state ${\vec{x}_{i}(k)}$ of $i$-th agent w.r.t
input ${\vec{u}_{i}(k)}$ are described by the following linear discrete-time systems:
\begin{equation}
\begin{matrix}
  \label{eq:systems}
\vec{x}_{i}(k+1)=A_{i}\vec{x}_{i}(k) + B_{i}\vec{u}_{i}(k)
\end{matrix}
\end{equation}

The $M$ subsystems are coupled under linear input constraints.
We assume as an interesting case when these constraints prevent the subsystems from meeting the systems' needs. Consequently, the constraints will always be active, yielding the same results from equality constraints~\cite{BoydVandenberghe2004}:
\begin{equation}
  \label{eq:constraint}
  \sum^{M}_{i=1}\Gamma_{i}\vec{u}_{i}(k)=\vec{u}_{\mathrm{\max}}
\end{equation}
where $\Gamma_{i}:\RR^{\pi_{\vec{u}_{i}(k)}\times \pi_{\vec{u}_{i}(k)}}$ and $\vec{u}_{\mathrm{\max}}:\RR^{\pi_{\vec{u}_{i}(k)}\times 1}$.

A known formulation of the MPC structure~\cite{CamachoBordons2007,VelardeEtAl2017,VelardeEtAl2017a,SimonEtAl2012,ChanfreutEtAl2018} with finite prediction horizon $N_{p}$ is the following:
\begin{problem}{Global MPC Problem.}\label{Pb:GOP}
\begin{equation*}
\begin{matrix}
\underset{\vec{u}_i(k:k+N_{p}-1|k)}{\mathrm{minimize}}&\resizebox{0.35\textwidth}{!}{$\overbrace{\sum\limits^{M}_{i=1} \overbrace{\sum_{j=1}^{N_{p}}\|\vec{v}_i(k+j|k)\|^{2}_{Q_i}+\|\vec{u}_i(k+j-1|k)\|^{2}_{R_i}}^{\textstyle J_{i}(k)}}^{\textstyle J_{G}(k)}$}\\
\mathrm{subject~ to}&~\eqref{eq:systems}\mathrm{~and~}\eqref{eq:constraint}
\left\}\small
\begin{aligned}
  &\forall i\in\{1,\dots,M\}\\
  &\forall j\in\{1,\dots,N_{p}\}
\end{aligned}\right.

\end{matrix}
\end{equation*}
with symmetric weight matrices ${Q_{i}\geq0}$, ${R_{i}>0}$. $\vec{v}_{i}(k)$ represents a control objective. It can either be ${\vec{v}_{i}(k)=\vec{w}_{i}(k)-\vec{x}_{i}(k)}$ for reference tracking, where $\vec{w}_{i}(k)$ is a state reference, or ${\vec{v}_{i}(k)=\vec{x}_{i}(k)}$ for disturbance rejection.

The optimal value of the problem~\ref{Pb:GOP}
is denoted by $J^{\star}$, and the optimal control sequences are represented by ${\vec{u}_i^{\star}(k:k+N_{p}-1|k)}$.
At each time $k$, the problem is solved, and the $\vec{u}_i^{\star}(k|k)$ are applied in each respective $i$ subsystem, following a receding horizon strategy.
\end{problem}

One can see that if the subsystems were not coupled by~\eqref{eq:constraint},  the overall system could be decomposed into $M$ parts, solvable in parallel.
Multiple decomposition methods solve this problem~\cite{BoydEtAl2011,PaulenEtAl2016,PflaumEtAl2014,ChanfreutEtAl2018}.
Still, since we are interested in resource constraints and the dual decomposition does not enforce local feasibility~\cite{BoydEtAl2015}, the primal decomposition is chosen.

\subsection{Distributed Model Predictive Control}\label{ssec:dMPC}

\newcommand{\masterpb}{\emph{master problem}}
The technique consists of decomposing the \emph{coupling constraints} (or \emph{complicating constraints}~\cite{BoydEtAl2015}) of the original optimization problem into local versions with additional variables
that are shared among them, negotiating the value of these variables until a consensus is reached.

Problem~\ref{Pb:GOP} is decomposed into multiple subproblems~\eqref{eq:DOP_local}, solvable in parallel, and a \masterpb~\eqref{eq:DOP_master}, which is equivalent to the original problem and uses information of the subproblems~\cite{BoydEtAl2015}:

  \begin{subequations}
    \begin{equation}
      \left.
        \small
        \begin{aligned}
          J_{i}^{\star}(\vec{\theta}_{i}(k))&=\underset{\vec{u}_{i}(k:k+N_{p}-1|k)}{\mathrm{minimize}}J_i(k)\\
          \mathrm{s.t.} &\quad\eqref{eq:systems}\\
          &\quad\Gamma_{i}\vec{u}_i(k)=\vec{\theta}_{i}(k):\vec{\lambda}_{i}(k)\\
        \end{aligned}
      \right\}
      \small
        \begin{aligned}
      &\forall i\in\{1,\dots,M\}\\
      &\forall j\in\{1,\dots,N_{p}\}
        \end{aligned}
      \label{eq:DOP_local}
    \end{equation}
    \begin{equation}
      \small
      \begin{aligned}
        J^{\star}&=\underset{\vec{\theta}(k:k+N_{p}-1|k)}{\mathrm{minimize}}\sum^{M}_{i=1} J^{\star}_i(\vec{\theta}_i(k))\\
        \mathrm{s.t.} &\quad \sum_{i=1}^{M}\vec{\theta}_{i}(k)=\vec{u}_{\max}
      \end{aligned}
      \label{eq:DOP_master}
    \end{equation}
  \end{subequations}

The subproblems~\eqref{eq:DOP_local} are formed by the local objectives $J_i(k)$  and a set of local constraints, with a sequence of allocations $\vec{\theta}_{i}(k:k+N_{p}-1|k)$ and associated sequence of dual variables (Lagrange multipliers) $\vec{\lambda}_{i}(k:k+N_{p}-1|k)$.
For brevity's sake, we drop the $(k:k+N_{p}-1|k)$ sequence notation, using only where pertinent.

The variables $\vec{\theta}_{i}$ represent the resource or the ``quantity'' allocated for each subproblem; thus, the names \mbox{``quantity decomposition''} and \mbox{``resource allocation''} are also given for this decomposition~\cite{Cohen1978}.

The \emph{master problem} shown in~\eqref{eq:DOP_master} can be solved using an iterative method that updates the allocation sequence $\vec{\theta}_{i}$.

Due to the form of the constraints, we use the projected sub-gradient method whose recurrence
equation is:
\begin{equation}
  \label{eq:projectedSubgradient}
\vec{\theta}^{(p+1)}=P_{\mathsf{H}}(\vec{\theta}\p-\rho\p\vec{g}\p)
\end{equation}
where ${\mathsf{H} = \enstq{\vec{\theta}}{\sum_{i=1}^{M}\vec{\theta}_{i}=\vec{u}_{\max}}}$, $\vec{g}\p$ is a sub-gradient of $J^{\star}(\vec{\theta}\p)$ at the instant $p$ and $\rho\p$ is an iteration step, well-chosen, so the method converges.

The sum $\sum_{i=1}^{M}\vec{\theta}_{i}$ can also be represented by the matrix multiplication $I_{c}^{M}\vec{\theta}$, where ${I_{c}^{M}=\1_{M,1}\otimes I_{c}}$. Where ${c=\pi_{\vec{u}_{i}(k:k+N_{p}-1|k)}=N_{p}\pi_{\vec{u}_{i}(k)}}$.

Assuming \emph{strong duality} holds, we can use the sensitivity analysis of the problem~\cite[\S~5.6.2]{BoydVandenberghe2004}, and we can conclude that the opposite of the sequences of optimal dual variables, $-\vec{\lambda}^{\star}_{i}$, which are $\vec{\theta}_{i}\p$ dependent, is a sub-gradient of $J_{i}^{\star}(\vec{\theta}_{i}\p)$, which can be used in~\eqref{eq:projectedSubgradient} to solve the problem~\eqref{eq:DOP_master}.

Applying the Euclidean projection onto $\mathsf{H}$~\cite{Ouyang2020} and using $-\vec{\lambda}_{i}^{\star}(\vec{\theta}_{i}\p)$ in~\eqref{eq:projectedSubgradient} results in the complete expression for the allocation's update~\cite[\S VI-C]{Cohen1978}:
\begin{equation}
  \label{eq:thetaNegot}
  \resizebox{0.45\textwidth}{!}{$
  \begin{aligned}
\vec{\theta}_{i}\pplusone=\vec{\theta}_{i}\p+\rho\left(\vec{\lambda}_{i}^{\star}(\vec{\theta}_{i}\p)-I_{c}^{M}{({I_{c}^{M}}\T I_{c}^{M})}^{-1}{I_{c}^{M}}\T \vec{\lambda}^{\star}(\vec{\theta}\p)\right)
  \end{aligned}$}
\end{equation}

In each step $(p)$, the subproblems receive a sequence of  allocation of the total resources.
Then they return their corresponding sequence of dual variables so the \emph{master problem} can be solved by updating the allocations, recommencing the negotiation.
Once a consensus is reached, the negotiation is finished, each subsystem takes the last sequence of inputs $\vec{u}_{i}^{\star}(k:k+N_{p}-1|k)$ calculated and applies the first element $\vec{u}_{i}^{\star}(k|k)$, following a receding horizon strategy.

\begin{figure}[!t]
  \centering
 \footnotesize \def\svgwidth{0.44\textwidth} 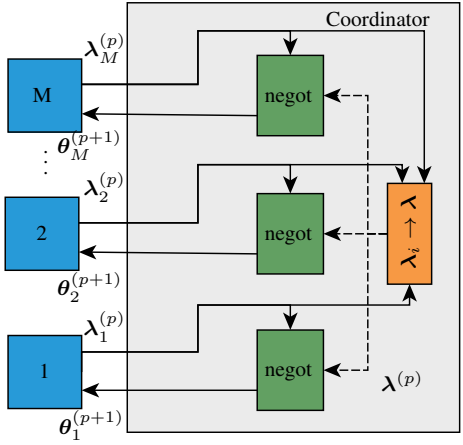
 \caption{Scheme of DMPC using a \emph{coordinator} and $M$ agents.}\label{fig:schemeQuantity}
\end{figure}

Delegating the iterative process of allocation update to an agent with the \emph{coordinator}'s role, we have the scheme in Fig.~\ref{fig:schemeQuantity} that illustrates the negotiation.
Observe that each block \emph{negot} solves~\eqref{eq:thetaNegot} for a respective agent $i$.
This way, the only interaction that the \emph{coordinator} has with the subsystems is via the variables $\vec{\lambda}_{i}\p$ and $\vec{\theta}_{i}\pplusone$, increasing the privacy of the subsystems.

Algorithm~\ref{alg:quantityAlg} resumes the distributed control problem solved to calculate the optimal input sequence at each time $k$ using quantity decomposition.
\begin{algorithm}[!t]
  \DontPrintSemicolon
  Coordinator initializes $\vec{\theta}^{(0)}$ \;
  $p:=0$\;
  \Repeat{$\|\vec{\theta}^{(p)} -\vec{\theta}^{(p-1)}\|\leq\epsilon$}{
  Subsystems solve~\eqref{eq:DOP_local}, and send $\vec{\lambda}^{\star}_{i}(\vec{\theta}\p)$\;
  Coordinator updates allocations~\eqref{eq:thetaNegot}\;
  $p:=p+1$
}
 \caption{Quantity decomposition based DMPC.}\label{alg:quantityAlg}
\end{algorithm}

\section{Attack in DMPC scheme}\label{sec:attack}

As expected~\cite{BoydEtAl2015,Cohen1978}, this decomposition method works well when each agent cooperatively calculates its $\vec{\lambda}_{i}$ correctly.
Here we study the effects when an ill-intentioned agent exploits the scheme for its interest.

\cite{VelardeEtAl2017,VelardeEtAl2017a,ChanfreutEtAl2018,AnandutaEtAl2020} present 4 types of attacks, which can be divided into 2 principal groups: changes in the optimization parameters (\emph{selfish attack} - multiply the objective function by a scalar~$\alpha$, \emph{fake reference}, and \emph{fake constraints}) and nonagreed control (\emph{liar agent}).
In the decomposition scheme used in this work, the coordinator allocates the resources.
So we can discard the last kind of attack.

Although we could make the same analysis from the mentioned works, we are interested in the coordinator's point of view, so any of these attacks will reflect as a change on the $\vec{\lambda}_{i}$ received.
Therefore, we propose that any selfish agent sends a
corrupted
\begin{equation}\label{eq:cheating}
\tilde{\vec{\lambda}}_{i}=\gamma_{i}(\vec{\lambda}_{i})
\end{equation}
to the coordinator instead of sending the agreed $\vec{\lambda}_{i}$.

We give a unidimensional example where ${\gamma_{i}(\lambda_{i})=\tau_{i}\,\lambda_{i}}$ to illustrate such an attack.
Here, 4 agents negotiate with the coordinator, and agent 1 attacks the system ($\tau_{1}\neq1$).

\begin{figure}[!t]
  \centering
  \includegraphics[width=\linewidth]{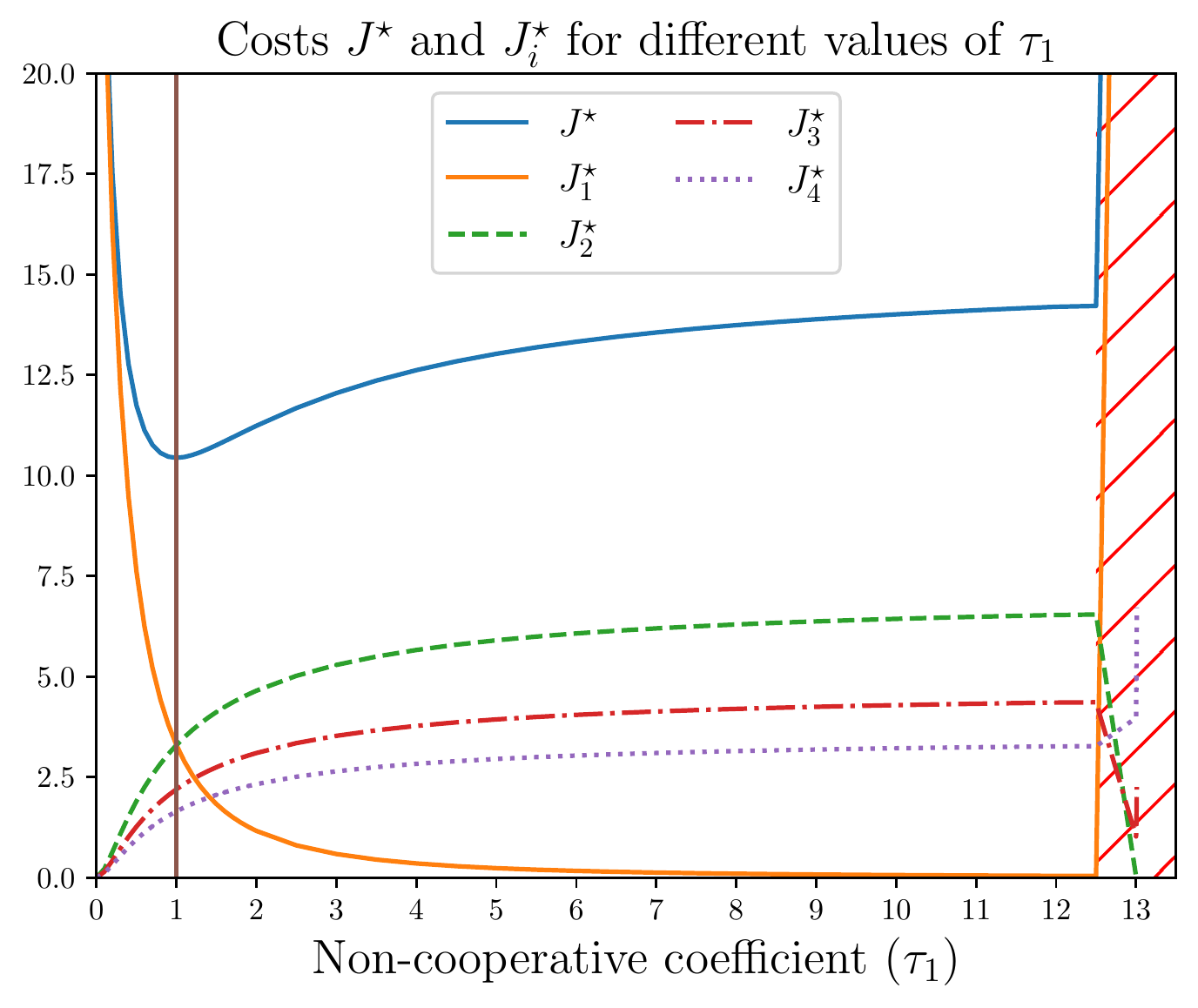}
  \caption{Change of $J^{\star}$ with respect to the non-cooperative coefficient $\tau_1$.}\label{fig:Jagainsttau}
\end{figure}

In Fig.~\ref{fig:Jagainsttau}, we see that when ${\tau_{i}>1}$, agents 1's local cost $J_{1}^{\star}$ decreases while all other costs, including the overall $J^{\star}$, increase.
This attack is comparable to the \emph{selfish attack} portrayed in~\cite{VelardeEtAl2017}.
This decrease in the cost justifies the attack since the attacking agent has more comfort than all others.

On the other hand, when $\tau_{1}$ tends to $0$, $J_{1}^{\star}$ increases and all others $J_{i}$ decreases, while still degrading the overall cost $J^{\star}$.
Such an agent could be considered as a benevolent agent or an agent attacked by a malevolent one.

From this variation in the values of $\vec{\lambda}_{1}$ caused by $\tau_{1}$, we can interpret its role in the negotiation: the values of $\vec{\lambda}_{i}$ represent the dissatisfaction with the given allocation $\vec{\theta}_{i}$.

Since the negotiation~\eqref{eq:thetaNegot} finds its stability when ${\vec{\lambda}\p-{I_{c}^M}{({I_{c}^M}\T {I_{c}^M})}^{-1}{I_{c}^M}\T\vec{\lambda} \p=\vec{0}}$, that means when all $\vec{\lambda}_{i}$ are equal to the mean of the $\vec{\lambda}_{i}$.
We can interpret that the coordinator's role is to minimize the overall dissatisfaction.
This way, the selfish agent can lie about its dissatisfaction (increasing $\vec{\lambda}_{i}$ by using an adequate $\gamma_{i}(\cdot)$), driving the negotiation to a value of $\vec{\theta}_{i}$ that ``satisfies'' it more (lower optimal value $J_{i}^{\star}$).

Another effect we can expect from the observation of~\eqref{eq:thetaNegot} is that the negotiation may not converge for some values of $\vec{\lambda}_{i}$. We can find those values by the analysis of the eigenvalues of the iterative process. This effect is illustrated in the hatched area in Fig.~\ref{fig:Jagainsttau}.

\section{Secure DMPC based on resource allocation}\label{sec:mitigation}

As seen, a malicious agent can deviate the allocations for its benefit, driving the negotiation or even destabilizing it.
Hence, it is needed to find a way to lessen the effects caused by this agent.
To fill this gap, we propose a detection and mitigation mechanism to reduce the effects of any agent malfeasance in the negotiation.
However, before presenting the mechanism, we need to analyze the problem structure to sustain the proposition.
\subsection{Quadratic Case --- Formal Analysis}\label{ssec:FA}
Another known form to represent the problems~\eqref{eq:DOP_local} is using matrix representation~\cite{ChanfreutEtAl2018}:
\begin{equation}
\begin{matrix}
\underset{\vec{U}_i(k)}{\mathrm{minimize}}& \overbrace{\frac{1}{2}{\vec{U}_i(k)}^T H_i\vec{U}_i(k)+{\vec{f}_i(k)}^T\vec{U}_i(k)}^{\textstyle J_{i}(\vec{\theta}_{i})}\\
\mathrm{s.t.}&\Theta_{i}\vec{U}_i(k)=\vec{\theta}_{i}:\vec{\lambda}_{i}\\
\end{matrix}
\label{eq:dmpcModLOP}
\end{equation}
If we take reference tracking, for instance, we have:
\begin{equation}
\small\begin{matrix*}[l]
 H_i&=&\mathcal{D}_i^T\bar{Q}_i\mathcal{D}_i+\bar{R}_i\\
\vec{f}_i(k)&=&\mathcal{D}_i^T\bar{Q}_i(\mathcal{M}_i\vec{x}_i(k)-\vec{W}_i(k))\\

\end{matrix*}
\label{eq:matrices}
\end{equation}
The input and setpoint predictions for times ${k}$ to ${k+N_{p}}$ calculated in time $k$ are adequately stacked in vectors $\vec{U}_{i}(k)$ and $\vec{W}_{i}(k)$.
${\mathcal{M}_{i}}$  and ${\mathcal{D}_{i}}$ are the prediction matrices of the MPC\@.
$\bar{Q}_{i}$, $\bar{R}_{i}$, and $\Theta_{i}$ are block diagonal matrices built repeating $N_{p}$ times $Q_{i}$, $R_{i}$, and $\Gamma_i$ respectively.

Notice that the matrices $H_{i}$ are not only symmetric positive definite, but they are also \mbox{time-invariant}, unlike the $\vec{f}_{i}(k)$, which depend on $x_{i}(k)$ and $\vec{W}_{i}(k)$.

Observe that since $J_{i}(\vec{\theta}_{i})$ is quadratic, we can get an explicit  solution for its dual variables $\vec{\lambda}_{i}$, which are affine with respect to $\vec{\theta}_{i}$:
\begin{equation}
  \begin{aligned}
    \label{eq:lambdafuntheta}
    \vec{\lambda}_{i}=-P_{i}\vec{\theta}_{i}-\vec{s}_{i}(k)
  \end{aligned}
\end{equation}
where ${P_{i}={(\Theta_{i}H_{i}^{-1}\Theta_{i}\T)}^{-1}}$ and ${\vec{s}_{i}(k)=P_{i}\Theta_{i}H_{i}^{-1}\vec{f}_{i}(k)}$.
We can observe that $P_{i}$ are symmetric and depend only on $\Theta_{i}$ and $H_{i}$, which are time-invariant.

\subsection{Detection and mitigation}
In this secure scheme, the exchange between coordinator and agents  is divided into two parts: first, to detect any misbehavior, and second, the negotiation itself, which limits the effects of eventual attacks.

\begin{assumption}
  $\gamma_{i}(\cdot)$ is the same during the negotiation phase for a given time $k$ (it does not depend on $p$).
\end{assumption}

\begin{assumption}
  We suppose the agent chooses a linear function such as
  \begin{equation}
    \label{eq:linear_cheating}
    \tilde{\vec{\lambda}_{i}}=\gamma_{i}(\vec{\lambda}_{i})=T_{i}(k)\vec{\lambda}_{i}
       =-T_{i}(k)P_{i}\vec{\theta}_{i}-T_{i}(k)\vec{s}_{i}(k),
  \end{equation}
and we define $\tilde{P}_{i}(k)=T_{i}(k)P_{i}$ and $\tilde{\vec{s}}_{i}(k)=T_{i}(k)\vec{s}_{i}(k)$.
\end{assumption}

Given that $P_{i}$ does not change from time to time, we can use the relation between $\vec{\theta}_{i}$ and $\vec{\lambda}_{i}$, shown in~\eqref{eq:lambdafuntheta}, to find estimates $\widehat{\tilde{P}}_{i}(k)$ such as:
\begin{equation}
  \label{eq:lambdafuntheta_tilde}
\tilde{\vec{\lambda}_{i}}=\gamma_{i}(\vec{\lambda}_{i}(\vec{\theta}_{i}))=-\widehat{\tilde{P}_{i}}(k)\vec{\theta}_{i}-\widehat{\tilde{\vec{s}}}_{i}(k)
\end{equation}

\begin{remark}
If the estimation does not converge, necessarily there has been a change in $P_{i}$ since the relation between $\vec{\lambda}_{i}$ and $\vec{\theta}_{i}$ has ceased to be affine.

\end{remark}

If we estimate $\widehat{\tilde{P}}_{i}(k)$ for two different times $k$ and they differ, then there has been a change in behavior in agent $i$.

\begin{assumption}\label{ass:Pnominal}
  We have access to the nominal value of $P_{i}$, denoted $\bar{P}_{i}$, from reliable attack-free historical data.
\end{assumption}

Using this strategy, we can detect a deviation from nominal behavior using ${E_{i}(k) =\|\widehat{\tilde{P}}_{i}(k)-\bar{P}_{i}\|_{F}}$, where ${\|\cdot\|_{F}}$ is the Frobenius norm.
Let ${d_{i}\in\{0,1\}}$ be an indicator that detects the attack in agent $i$.
If the disturbance $E_{i}(k)$ respects an arbitrary bound
\begin{equation}
  \label{eq:2}
  E_{i}(k)\leq\epsilon_{P},
\end{equation}
then  ${d_{i}=0}$, and no attack is detected. Otherwise, ${d_{i}=1}$, and a change in behavior of agent $i$ is detected.

If the attack is detected and we want to counteract the change in $\vec{\lambda}_{i}$, one strategy would be to recover $\vec{\lambda}_{i}$ from an inverse of $\gamma_{i}(\cdot)$.

\begin{assumption}
We suppose ${\tilde{\vec{\lambda}}_{i}=\vec{0}}$ only if ${\vec{\lambda}_{i}=\vec{0}}$, which implies $T_{i}(k)$ invertible.
\end{assumption}

Using these assumptions, we can try to estimate the inverse of $T_{i}(k)$ as in  \begin{equation}\label{eq:5}
{\widehat{T_{i}(k)^{-1}}=\bar{P}_{i}\widehat{\tilde{P}_{i}}(k)^{-1}},
\end{equation}
and from~\eqref{eq:lambdafuntheta}, we can derive a method to reconstruct $\vec{\lambda}_{i}$:
\begin{equation}
  \label{eq:lambdareconstruction}
  {\vec{\lambda}_{i}}_{\mathrm{rec}}=\widehat{T_{i}(k)^{-1}} \tilde{\vec{\lambda}_{i}} =-\bar{P}_{i}\vec{\theta}_{i}-\widehat{{T_{i}(k)^{-1}}}\widehat{\tilde{\vec{s}}}_{i}(k).
\end{equation}
Notice that we also need $\widehat{\tilde{\vec{s}}}_{i}(k)$ to use this reconstruction.

This reconstructed ${\vec{\lambda}_{i}}_{\mathrm{rec}}$ can be used in~\eqref{eq:thetaNegot}. Observe that, as~\eqref{eq:lambdareconstruction} does not depend on $\tilde{\vec{\lambda}}_{i}$, the rest of the negotiation process takes place without taking the attacking agent's responses into account.

In case no attack is detected, the coordinator can use the $\tilde{\vec{\lambda}_{i}}$ during the negotiation phase.

\begin{figure}[b]
  \centering
  \scriptsize \def\svgwidth{0.49\textwidth}
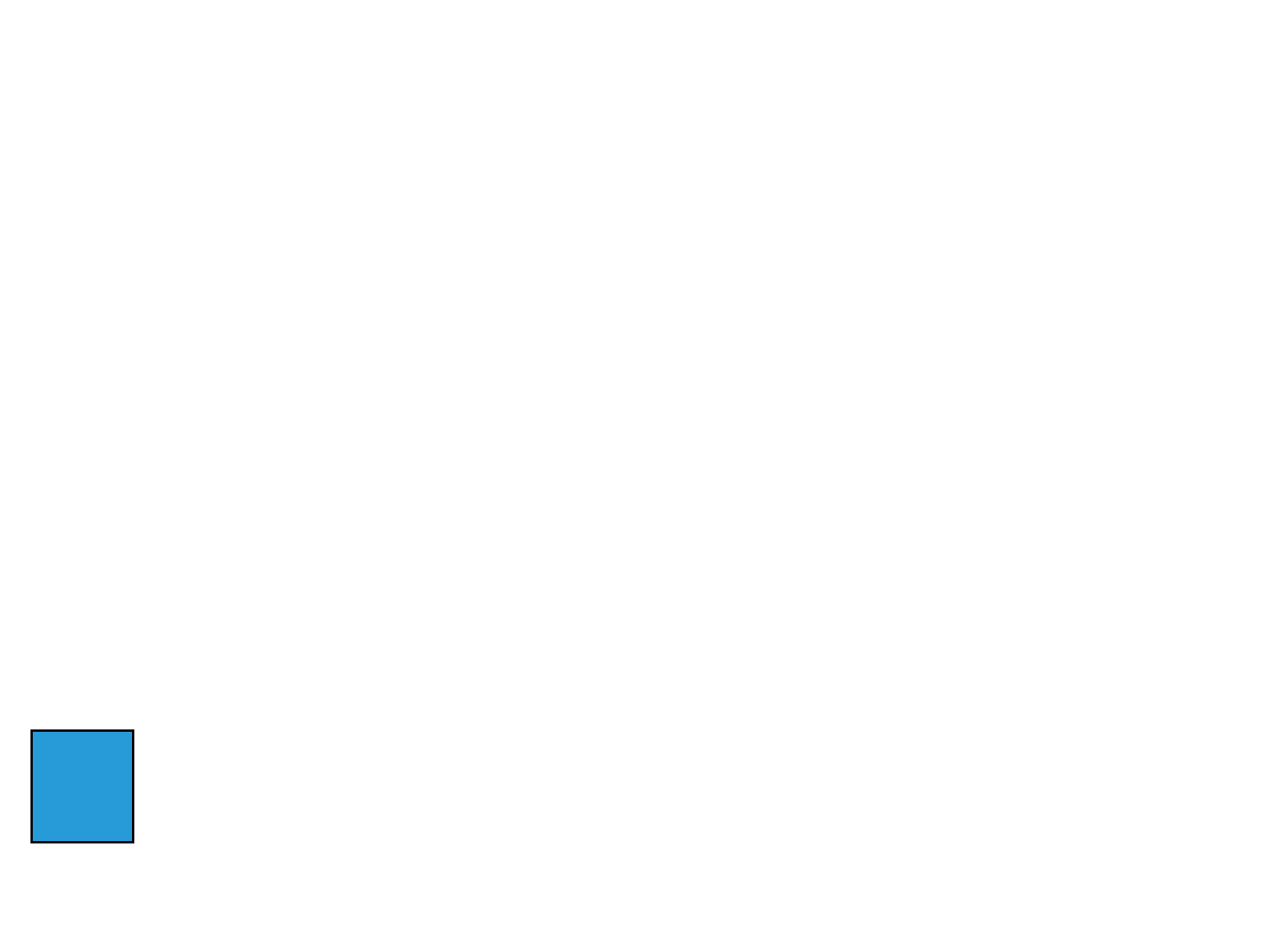
  \caption{Scheme for Secure DMPC}\label{fig:schemeSafeQuantity}
\end{figure}

This mechanic of detecting and choosing which version of $\vec{\lambda}$ to use during the negotiation, corresponds to the inclusion of a supervisor for each agent (Fig.~\ref{fig:schemeSafeQuantity}),

Observe in Fig.~\ref{fig:schemeSafeQuantity} that the coordinator sends $\mathring{\vec{\theta}}_{i}$ to the agents. These values may be the ones from the negotiation or other. The reason to send different values is discussed in the following subsection.

\subsection{Considerations about parameter estimation}

As seen, we need to estimate $\tilde{P}_{i}(k)$ and $\tilde{\vec{s}}_{i}(k)$.
This estimation is achieved by the relation between $\vec{\theta}_{i}$ and $\vec{\lambda}_{i}$ shown in~\eqref{eq:lambdafuntheta}.
As we suppose there is no noise in the communication, we propose to use Recursive Least Squares (RLS) with a forgetting coefficient $\phi$ to find simultaneously unbiased estimates of $\tilde{P}_{i}(k)$ and $\tilde{\vec{s}}_{i}(k)$.

If we try to estimate during the negotiation, the estimation will fail since consecutive values of $\vec{\lambda}_{i}^{p}$ and $\vec{\theta}_{i}^{p}$ are necessarily linearly dependent~\eqref{eq:thetaNegot}, and estimators become badly scaled. This fact is known and is described as low input excitation~\cite[\S 5]{AastroemWittenmark1989}.
As a counter-measure, to enrich the input excitation, the coordinator sends a sequence of random values of $\vec{\theta}_{i}$ until the estimation converges. It then resumes the typical negotiation, eventually using the mitigation mechanism if an attack is detected.

\begin{assumption}\label{ass:tildePinvertible}
Since $P_{i}$ is expected to be symmetric~\eqref{eq:lambdafuntheta}, we suppose that the attacker chooses a $T_{i}(k)$ that does not change the structure of the resulting matrix, so it can not be discovered.
In this case, we assume $\tilde{P}_{i}(k)$ symmetric and invertible.
\end{assumption}
As $\tilde{P}_{i}(k)$ is symmetric, we estimate only the upper triangle, reducing the number of estimated parameters from $\pi_{\tilde{P}_{i}(k)}$ to $\frac{\pi_{\tilde{P}_{i}(k)}+\sqrt{\pi_{\tilde{P}_{i}(k)}}}{2}$, and consequently the length of the estimation sequence~\cite{AastroemWittenmark1989}.

We stack the elements of $\widehat{\tilde{P}}_{i}(k)$ and $\widehat{\tilde{\vec{s}}}_{i}(k)$ estimated in a step $h$ in vectors $\eta_{i}^{h}$. The estimation converges when
${\|\eta_{i}^{h}-\eta_{i}^{h-1}\|\leq\epsilon}$,
with $\epsilon$ arbitrarily small.

\subsection{Secure DMPC}
\label{sec:secure-dmpc}
After all the reflections about parameter estimation and the detection and mitigation mechanism, we can finally propose a secure DMPC based on the reconstruction of $\vec{\lambda}_{i}$.

Algorithm~\ref{alg:safeDMPC} summarizes the process used to find the optimal inputs $\vec{u}_{i}^{\star}(k|k)$ to be applied at each time $k$.
We can see the two phases: the detection phase, where the coordinator detects if the system is attacked and by which agent. And the second phase, where the usual negotiation in algorithm~\ref{alg:quantityAlg} takes place, using different values of $\vec{\lambda}_{i}$ depending on if the respective agent is an attacker.

In the next section, we present an example to illustrate the performance of the mechanism.
\SetKwBlock{negotPhase}{ Negotiation Phase:}{}
\SetKwBlock{detectPhase}{ Detection Phase:}{}
\begin{algorithm}[h]
  \DontPrintSemicolon
  \detectPhase{
  $h:=0$\;
    \Repeat{
$\|\eta_{i}^{h}-\eta^{h-1}\|\leq\epsilon$
}{
    Coordinator sets random $\vec{\theta}_{i}^{(h+1)}$ \;
    Subsystems solve~\eqref{eq:DOP_local}, and send $\vec{\lambda}^{\star}_{i}(\vec{\theta}^{(h)})$\;
    Coordinator estimates $\widehat{\tilde{P}}_{i}(k)^{(h)}$ and $\widehat{\tilde{\vec{s}}}_{i}(k)^{(h)}$ \;
    $h:=h+1$\;
    }
    Coordinator computes $d_{i}$ using~\eqref{eq:2}\;
  }
  \negotPhase{
  Coordinator initializes $\vec{\theta}^{(0)}$ \;
  $p:=0$\;

  \Repeat{$\|\vec{\theta}^{(p)} -\vec{\theta}^{(p-1)}\|\leq\epsilon$}{
  Subsystems solve~\eqref{eq:DOP_local}, and send $\vec{\lambda}^{\star}_{i}(\vec{\theta}\p)$\;

  Coordinator updates allocation~\eqref{eq:thetaNegot} using adequate versions of $\vec{\lambda}_{i}$ for each agent: $\vec{\lambda}_{i}^{\star}(\vec{\theta}\p)$, if $d_{i}=0$ and ${\vec{\lambda}_{i}}_{\mathrm{rec}}$, if ${d_{i}=1}$\;
  $p:=p+1$
 }}
 \caption{Secure DMPC.}\label{alg:safeDMPC}
\end{algorithm}

\section{Example: Temperature Control}\label{sec:App_dMPC}
In this example, we want to control the temperature of 4~distinct~rooms (called I, II, III, and IV) under power scarcity using quantity decomposition.
The systems are modeled as continuous-time linear \mbox{time-invariant} systems using the \mbox{3R-2C} model~\cite{GoudaEtAl2002}.

The state-space model of each subsystem is given by:
\begin{equation}
\begin{matrix}
  \dot{\left[\begin{matrix} {\vec{x}_{A}}_{i}\\ {\vec{x}_{W}}_{i} \end{matrix} \right]}
  =\dot{\vec{x}}_{i}=A_{\mathrm{c}_{i}}\vec{x}_{i}+B_{\mathrm{c}_{i}}\vec{u}_{i}\\
\vec{y}_{i}=C_{\mathrm{c}_{i}}\vec{x}_{i}
\end{matrix}
\end{equation}
where
\begin{equation}
  \label{eq:4}
  \begin{matrix}
  A_{\mathrm{c}_{i}}=\left[
    \begin{matrix}
      -\frac{1}{C_{\mathrm{res}_{i}}Rf_{i}}-\frac{1}{C_{\mathrm{res}_{i}}Ri_{i}}& \frac{1}{C_{\mathrm{res}_{i}}Ri_{i}}\\
      \frac{1}{Cs_{i}Ri_{i}} &-\frac{1}{Cs_{i}Ro_{i}}-\frac{1}{Cs_{i}Ri_{i}}
    \end{matrix}\right]\\
  \begin{matrix}
    B_{\mathrm{c}_{i}}=\left[
      \begin{matrix}  \frac{10}{C_{\mathrm{res}_{i}}}& 0\end{matrix}
    \right]\T&C_{\mathrm{c}_{i}}=\left[\begin{matrix}1 & 0\end{matrix}\right]
  \end{matrix}
  \end{matrix}
\end{equation}

We can see the meaning and the values of its parameters in tables~\ref{tab:modelParamMeaning} and~\ref{tab:modelParam}.

\begin{table}[b]
  \centering
  \caption{Model Parameters Meanings}\label{tab:modelParamMeaning}
  \begin{tabular}[b]{cl}
    \toprule
    Symbol&Meaning\\
    \midrule
    $C_{\mathrm{res}_{i}}$&Heat Capacity of Inside Air\\
    $Cs_{i}$&Heat Capacity of External Walls\\
    $Rf_{i}$&Resistance Between Inside and Outside Air (from windows)\\
    $Ri_{i}$&Resistance Between Inside Air and Inside Walls\\
    $Ro_{i}$&Resistance Between Outside Air and Outside Walls\\
    \bottomrule
  \end{tabular}
\end{table}
\begin{table}[b]
  \centering
  \caption{Model Parameters Values}\label{tab:modelParam}
  \begin{tabular}[t]{cccccc} \toprule
    Symbol& I & II & III & IV &Unit\\
    \midrule
    $C_{\mathrm{res}}$    &$5$   & $4$ & $4.5$ &$4.7$ &$10^{4}\mathrm{J/K}$ \\
    $Cs$               &$8$   & $7$ &$9$&$6$  &$10^{4}\mathrm{J/K}$  \\
    $Rf$               &$5$   & $6$& $4$& $5$  &$10^{-3}\mathrm{K/W}$ \\
    $Ri$               &$2.5$ & $2.3$&$2$&$2.2$ &$10^{-4}\mathrm{K/W}$\\
    $Ro$               &$0.5$ & $1$ &$0.8$&$0.9$  & $10^{-4}\mathrm{K/W}$ \\
    \bottomrule
  \end{tabular}
\end{table}
The states ${\vec{x}_{A}}_{i}$ and ${\vec{x}_{W}}_{i}$ represent the mean temperatures of the air and walls inside room $i$. The input $\vec{u}_{i}$ is the heating power
for the corresponding room.
The global coupling constraint is ${\sum_{i=1}^{4}\vec{u}_{i}(k)=4\mathrm{kW}}$.

The subsystems are discretized using the zero-order hold discretization method with sampling time
${T_{s}=0.25\mathrm{h}}$
and the quantity decomposition-based DMPC is implemented using prediction horizon ${N_{p}=4}$.

Three scenarios are simulated for a period of 5 hours:

  \begin{enumerate}
    \item Nominal behavior.
    \item \mbox{Agent I presents constant non-cooperative behavior} ${T_{I}(k)=4\,I_{\sqrt{\pi_{P_{I}}}}}$ for ${k\geq6}$, without correction.
    \item \mbox{Agent I presents constant non-cooperative behavior} ${T_{I}(k)=4\,I_{\sqrt{\pi_{P_{I}}}}}$ for ${k\geq6}$, with correction, ${\epsilon_{P}=10^{-4}}$.
  \end{enumerate}

\begin{figure}[t]
  \centering
  \includegraphics[width=7cm]{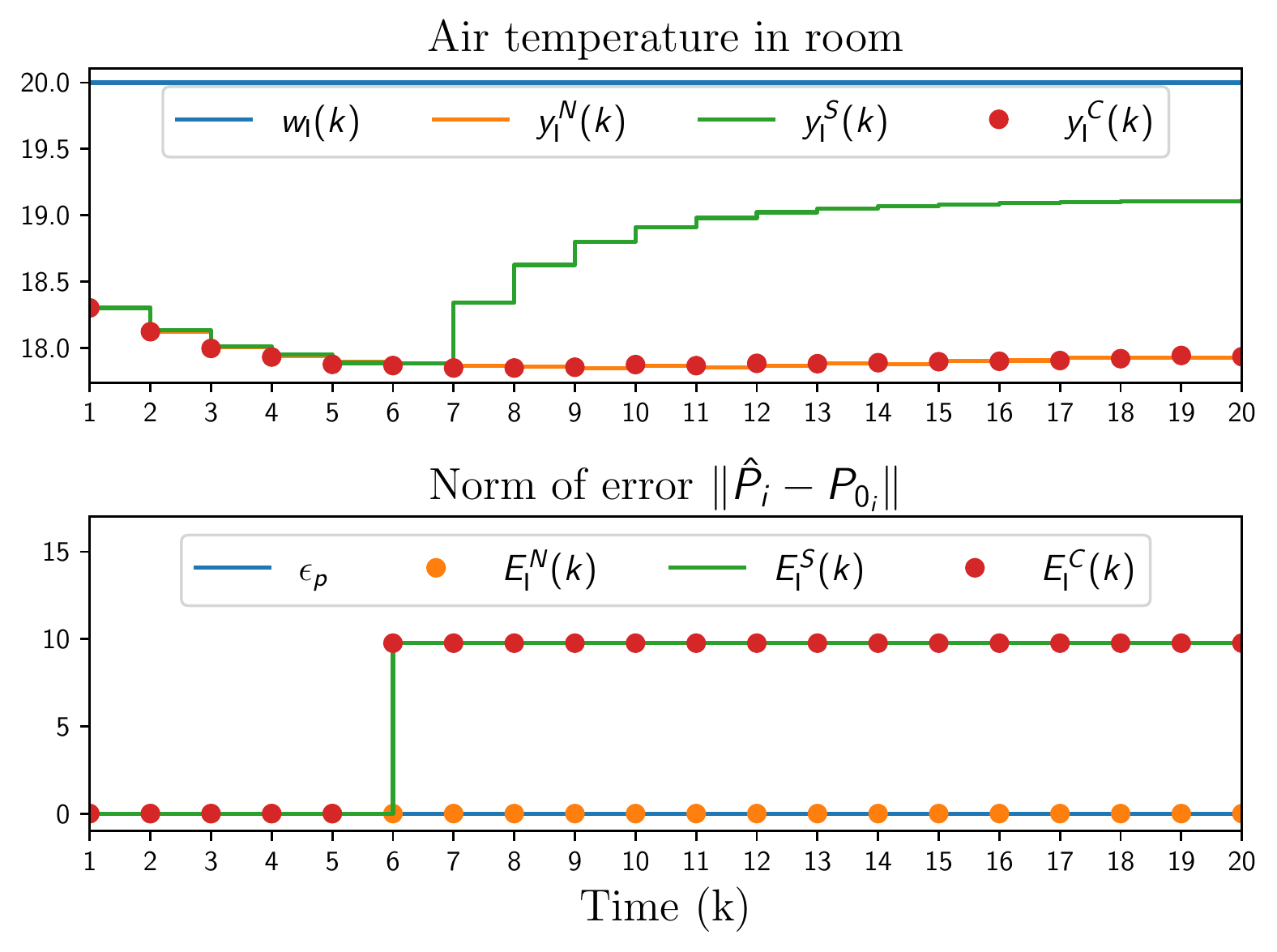}
\caption{Air temperature in room I and the decision variable $E_{I}(k)$ for
different scenarios: nominal (N), with selflish behavior without correction (S)
and with selfish behavior with correction (C)}\label{fig:response3Scenarios}
\end{figure}

In Fig.~\ref{fig:response3Scenarios}, first, we compare the output of the agent I
(air temperature in the room) with its reference ($20^{\circ}$C), and then the
decision variable ${E_{I}(k)}$ with the threshold $\epsilon_{P}$.
All the 3 scenarios above are represented with indices N (for nominal), S (for selfish), and C (for corrected).

Observe that in the nominal behavior, the reference $w_{I}$ is not reached
due to power scarcity since we deliberately set a total power not sufficient to
satisfy the needs of each agent.
As expected, the decision variable lies under the threshold $\epsilon_{P}=10^{-4}$ with values of order ${E_{I}^{N}(k)\approx10^{-10}}$.

When the agent presents a selfish behavior, the tracking error
${w_{I}-y_{I}}$ is reduced but insufficient to attain the reference.
In this case, the detection variable surpasses $\epsilon_{P}$,
${E_{I}^{S}=9.762}$, indicating the change of behavior of agent I.

When the correction is activated in the system, we see that the corrected
$y_{I}^{C}$ approaches the nominal value $y_{I}^{N}$, illustrating the good
performances of our proposition.

We can also evaluate the performance of the proposed mechanism by comparing the
local and global costs calculated using the initial cost function presented
in~\eqref{eq:dmpcModLOP} using N as the total period of simulation, ${N=20}$.
The same 3 scenarios are compared in table~\ref{tab:costsGlobalLocal}.

As in Section~\ref{sec:attack}, when agent I is selfish, we see the decline of its cost at the expense of increasing all other costs. This increase in cost degrades the global objective.
When the correction mechanism is activated, the differences between costs are minimal, and the global cost stays close to the nominal value, highlighting the mechanism's performance.
\begin{table}[t]
  \centering
  \caption{Comparison of costs $J_{i}^{N}$ and $J_{G}^{N}$}\label{tab:costsGlobalLocal}
  \begin{tabular}[t]{cccc} \toprule
    Agent  & Nominal & Selfish & Selfish + correction\\
    \midrule
    I      & 103  &  64  & 104  \\
    II     &  73  &  91  &  73  \\
    III    & 100  & 123  & 101  \\
    IV     & 132  & 154  & 131  \\
    Global & 408  & 442  & 409  \\
    \bottomrule
  \end{tabular}
\end{table}
\enlargethispage{-.8cm}

\section{CONCLUSION AND FUTURE WORKS}\label{sec:CC}

In this paper, an algorithm for monitoring and correcting exchanges between agents in a resource-sharing system has been proposed.
The algorithm exploits the particular structure of exchanges, part of which must be constant over time.
The first phase consists of identifying this constant part and checking if an attacker has modified it.
From this identification, it is possible to reconstruct the original mechanism and find the centralized optimality.
This principle should be generalized to other types of decomposition structures, and this is what we plan to do in the near future.

\section{ACKNOWLEDGMENTS}\label{sec:ACK}

The authors would like to acknowledge C. R. Sorgho for her preliminary results.

\bibliographystyle{IEEEtran}
\bibliography{bibliography}

\end{document}